\documentclass[preprint,secnumarabic,amssymb,nobibnotes,nofootinbib,aps,prc]{revtex4}
\usepackage{psfig}
\usepackage{graphicx}
\usepackage{bm}
\newcommand{\bea}{\begin{eqnarray}}
\newcommand{\eea}{\end{eqnarray}}
\newcommand{\be}{\begin{equation}}
\newcommand{\ee}{\end{equation}}
\newcommand{\bt}{\begin{tabular}}
\newcommand{\et}{\end{tabular}}

\newcommand{\beas}{\begin{eqnarray*}}
\newcommand{\eeas}{\end{eqnarray*}}

\begin{document}
{\hspace{\fill} TUM/T39-03-31}
\baselineskip 6.5mm
\date{\today}
\title{Dilepton emission rates from hot hadronic matter}
\author{Thorsten Renk}
\email{trenk@physik.tu-muenchen.de}
\affiliation{Physik Department, Technische Universit\"at M\"unchen,
D - 85745, Garching, Germany}

\author{Amruta Mishra}
\email{mishra@th.physik.uni-frankfurt.de}
\affiliation{Institut f\"ur Theoretische Physik,
        Robert Mayer Str. 8-10, D-60054 Frankfurt am Main, Germany}

\date{\today} 

\begin{abstract}
The vacuum polarisation effects from the nucleon sector lead to large
medium modifications of the vector meson masses in the Walecka model.
With a quantum hadrodynamic framework including the quantum effects, 
and using a quasiparticle description for quark gluon plasma(QGP), 
the dilepton emission rate from the hot and dense matter resulting 
from relativistic nuclear collisions is calculated.
Using a model for the fireball
evolution which has been shown to reproduce other observables
such as charmonium suppression, photon emission and the abundance of
hadronic species, we compare with the dilepton invariant mass spectrum
measured by the CERES collaboration at CERN SPS. We also compare
the results to a previous calculation where the spectral function of
a virtual photon (a key ingredient for the emission rate) has been
calculated in a chiral SU(3) model.
\end{abstract}
\draft
\maketitle

\def\bfm#1{\mbox{\boldmath $#1$}}

\section{Introduction}

The study of the in-medium properties of the vector mesons 
($\rho$ and $\omega$) in hot and dense matter is actively
investigated, both experimentally 
\cite {expt,rhic} and theoretically \cite{brown,hat}. 
The experimental observation of enhanced dilepton production
\cite{expt} in the low invariant mass regime possibly is due 
to a reduction in the vector meson masses in the medium. 
It was first suggested by Brown and Rho that the vector meson masses 
drop in the medium according to a simple scaling law \cite{brown}. 
Within the  framework of Quantum Hadrodynamics (QHD),
the vacuum polarisation effects from the
baryon sector \cite{hatsuda,hatsuda1,jeans,sourav} lead to a significant
drop of the vector meson masses in the medium whereas the mass modification
is marginal with only Fermi sea polarisation effects.
The properties of the hadrons as modified in the thermal bath
are reflected in the dilepton and photon spectra emitted
from a hot and dense matter \cite{sourav,wambach,brat,koch}.
Dileptons are interesting probes for the study of the hot and dense 
matter formed in relativistic heavy ion collisions. 
Since they do not interact strongly, the dileptons escape
unthermalized from the hot and dense matter at all stages of the
evolution. The observed enhancement of dileptons in the
low invariant mass regime has initiated extensive theoretical 
investigations on the temperature \cite{temp} and density \cite{dens}
modifications of the dileptons from hot hadronic matter as well as
from a quark gluon plasma (QGP) formed in heavy ion collisions.
The increased dilepton yield in S+Au collisions as observed by
the CERES collaboration was attributed to enhanced $\rho$-meson
production via $\pi^+ \pi^-$ annihilation and a dropping of the $\rho$-mass
in the medium \cite{likobr,elina}. A large broadening of the $\rho$-meson
spectral function arising due to scattering off by baryons 
\cite{rapp,chiral} has been shown to reproduce the CERES data quite well.
In the present investigation, the space--time evolution of the strongly 
compressed hadronic matter formed in a relativistic heavy ion collision 
is considered in a mixed scenario of QGP and hadronic matter.
The QGP is described within a quasiparticle
picture \cite{quasi} and the hadronic matter described in a Quantum 
Hadrodynamic framework including quantum fluctuation effects from the
baryon and scalar meson sectors \cite{mishra,vecmass,dlp}.
The dilepton emission rates from the expanding fireball are then studied
in a model based on local thermal equilibrium and isentropic expansion
\cite{fireball}.

It was earlier demonstrated in a nonperturbative formalism that a
realignment of the ground state with baryon-antibaryon condensates is
equivalent to the relativistic Hartree approximation (RHA)
\cite {mishra}. The ground state for the nuclear matter was extended
to include sigma condensates to take into account the quantum
correction effects from the scalar meson sector \cite {mishra}. 
Such a formalism includes multiloop effects and is self consistent
\cite{mishra,pi}. The methodology 
was then generalized to consider hot nuclear matter \cite{hotnm} as
well as to study hyperon-rich dense matter \cite{shm}
relevant for neutron stars. The effect of vacuum polarisations on the
vector meson properties \cite{vecmass} and on the
static dilepton spectra \cite{dlp} has also been recently studied.
In the low invariant mass regime, the scalar meson
contributions lead to considerable broadening the $\omega$ peak
in the dilepton spectra as compared to RHA, which leads to smearing
and ultimate disappearance of the $\omega$ peak 
at high densities \cite{dlp}. The present investigation 
of dilepton spectra using a dynamical fireball model 
is compared with results obtained using other descriptions 
for hadronic matter \cite{chiral,likobr}, 
as well as to the experimental results from CERES collaboration \cite{expt}. 

We organize the paper as follows. We first briefly recapitulate 
the nonperturbative framework used for studying the medium modification
of the $\omega$ and $\rho$ vector meson properties
including the vacuum fluctuation effects in the Walecka model,
in section 2. 
Section 3 discusses the parametrisation of the photon self energy
in the hadronic phase. The quark gluon plasma (QGP) phase is treated
within a quasiparticle picture \cite{quasi} as described in section 4. 
In section 5, we describe briefly the model considered for the dynamical
evolution of the fireball. Section 6 deals with the calculation of dilepton
emission from the fireball. Finally, we discuss the findings of the
present investigation in section 7 and summarise in section 8.

\section {Vacuum polarisation effects and in-medium vector meson properties}

We briefly recapitulate here the vacuum polarisation effects arising
from the nucleon and scalar meson fields in hot nuclear matter in a
nonperturbative variational framework \cite{hotnm} and their influence 
on the vector meson properties in the hot and dense matter. 
The method of thermofield dynamics (TFD) \cite {tfd} is used here to
study the ``ground state" (the state with minimum thermodynamic potential)
at finite temperature and density within the Walecka model with a quartic
scalar self interaction. The temperature and density dependent baryon
and sigma masses are also calculated in a self-consistent manner in
the formalism. The ansatz functions involved in such an approach are
determined through functional minimisation of the thermodynamic
potential.

The Lagrangian density in the Walecka model is given as
\begin{eqnarray}
{\cal L}&=&\bar \psi \left(i\gamma^\mu \partial_\mu
-M-g_\sigma \sigma-g_\omega\gamma^\mu \omega_\mu\right)\psi
+\frac{1}{2}\partial^\mu\sigma
\partial_\mu\sigma-\frac{1}{2} m_\sigma ^2 \sigma^2
-\lambda \sigma^4
\nonumber\\
&& +\frac{1}{2} m_\omega^2 \omega^\mu \omega_\mu
-\frac{1}{4}(\partial_\mu \omega_\nu -\partial_\nu \omega_\mu)
(\partial^\mu \omega^\nu -\partial^\nu \omega^\mu).
\end{eqnarray}
In the above, $\psi$, $\sigma$, and $\omega_\mu$ are the fields for
the nucleon, $\sigma$, and $\omega$ mesons with masses $M$,
$m_\sigma$, and $m_\omega$ respectively. 
The quartic coupling term in $\sigma$ is necessary for the sigma
condensates to exist, through a vacuum realignment \cite{mishra}.  Our
calculations thus include the quantum effects arising from the sigma
meson in addition to the mean field contribution from the the quartic
self interaction of the scalar meson.  We retain the quantum nature of
both the nucleon and the scalar meson fields, whereas the vector
$\omega$-- meson is treated as a classical field, which in the mean field
approximation is given as $\langle \omega^\mu
\rangle=\delta_{\mu 0} \omega_0$. The reason is that without
higher--order term for the $\omega$-meson, the
quantum effects generated due to the $\omega$-meson through the
present variational ansatz turn out to be zero.
The thermodynamic quantities including the quantum effects
can then be written down. The details regarding the formalism can
be found in earlier references \cite {mishra,hotnm,vecmass}.
Extremisation of the thermodynamic potential, with respect to 
the meson fields $\sigma_0$ and $\omega_0$ yields the self--consistency 
conditions for $\sigma_0$ (and hence for the effective nucleon mass, 
$M^*=M+g_\sigma \sigma_0$), and for the vector meson field $\omega_0$.
We might note here that the quantum effects arising from the
scalar meson sector through $\sigma$ meson condensates amount to a
sum over a class of multiloop diagrams and, do not correspond to the
one meson loop approximation for scalar meson quantum effects
considered earlier \cite{serot}.  

\subsection {In-medium vector meson masses}

We now examine the medium modification to the masses 
of the $\omega$- and $\rho$-mesons in hot nuclear matter 
including the quantum correction effects 
in the relativistic random phase approximation. 
The interaction vertices for these mesons with nucleons are given as

\begin{equation}
{\cal L}_{\rm int}=g_{V}\Big (\bar \psi \gamma_\mu \tau^a \psi V_a^{\mu}
-\frac {\kappa_V}{2 M_N} \bar \psi \sigma_{\mu \nu} \tau^a \psi
\partial ^\nu V_a^\mu  \Big)
\label{lint}
\end{equation}
\noindent where $V_a^\mu=\omega^\mu$ or $\rho _a^\mu $, $M_N$ is the free
nucleon mass, $\psi$ is the nucleon field and $\tau_a=1$ or $\vec \tau $,
$\vec \tau$ being the Pauli matrices. $g_V$ and $\kappa_V$ correspond
to the couplings due to the vector and tensor interactions for the
corresponding vector mesons to the nucleon fields.  
The vector meson self energy is expressed 
in terms of the nucleon propagator, $G(k)$ modified by the quantum effects.
This is given as
\begin{equation}
\Pi ^{\mu \nu} (k)=-\gamma_I g_V^2 \frac {i}{(2\pi)^4}\int d^4 p\, 
{\rm Tr} \Big [ \Gamma_V^\mu (k) G(p) \Gamma_V^\nu (-k) G(p+k)\Big],
\end{equation}
where $\gamma_I=2$ is the isospin degeneracy factor for
nuclear matter, and $\Gamma_V^\mu (k)=\gamma^\mu \tau_a -
\frac {\kappa_V}{ 2 M_N}\sigma^{\mu \nu}$
represents the meson-nucleon vertex function. For the $\omega$
meson, the tensor coupling, being small as compared to the
vector coupling to the nucleons \cite {hatsuda1}, is 
neglected \cite{vecmass,dlp}. 
After carrying out the renormalization procedures for the vector 
self energies, the effective mass of the vector meson is obtained 
by solving the equation
\begin{equation}
k_0^2-m_V^2 + {\rm Re} \Pi (k_0,{\bfm k}=0) =0.
\label {omgrho}
\end{equation}
where $\Pi =\frac {1}{3} \Pi^\mu _\mu$. 

\subsection {Meson decay properties}

For a baryon-free environment, $\rho \rightarrow \pi^+ \pi^-$ is
the dominant decay channel for $\rho$ meson.
The decay width for this process is
calculated from the imaginary part of the self energy using the
Cutkosky rule, and in the rest frame of the $\rho$-meson 
is given by
\begin{equation}
\Gamma _\rho (k_0)=\frac {g_{\rho \pi \pi}^2}{48\pi} 
\frac {(k_0^2-4 m_\pi^2)^{3/2}}{k_0^2} \Bigg [
\Big (1+f(\frac {k_0}{2}) \Big )
\Big (1+f(\frac {k_0}{2}) \Big )
-f(\frac {k_0}{2}) f(\frac {k_0}{2}) \Bigg ]
\label {gmrho}
\end{equation}
where, $f(x)=[e^{\beta x}-1]^{-1}$ is the Bose-Einstein distribution 
function. The first and the second terms in the equation (\ref{gmrho})
correspond to the decay and the formation of the resonance, $\rho$.
In the calculation
for the $\rho$ decay width, the pion has been treated as free, 
and any modification of the pion propagator due to 
effects like delta-nucleon hole excitation \cite {asakawa} 
to yield a finite decay width for the pion, have not
been taken into account. 
The coupling $g_{\rho \pi \pi}$ is fixed from the decay width
of $\rho$ meson in vacuum ($\Gamma_\rho$=151 MeV) decaying 
to two pions.

In the presence of baryons, however, it has been shown that there 
is considerable increase of the $\rho$ decay width
\cite{chiral} in the thermal medium due to the
scattering off the baryons. The dominant contributions
which lead to appreciable broadening of the $\rho$- spectral function
are the inelastic processes $\rho N \rightarrow \pi N$
and $\rho N \rightarrow \pi \Delta$. 
The imaginary parts of the corresponding scattering amplitudes
in the large baryon mass limit are given as \cite{chiral}
\be
{\rm Im T} _{\rho N}^{(\pi N)}= g_A^2 {\cal H} (k_0,0,m_\pi)
\label{pinn}
\ee
and
\be
{\rm Im T} _{\rho N}^{(\pi \Delta)}=2 g_A^2
 {\cal H} (k_0,m_\Delta ^*-m_N^*,m_\pi)
\label{piddel}
\ee
 where,
\begin{eqnarray}
{\cal H}(k_0,\Delta,m) &= & \frac {g_{\rho NN}^2}{6 \pi f_\pi^2}
\Big [ \frac {\sqrt {(k_0-\Delta)^2 -m^2}}{(k_0^2-2 k_0 \Delta)^2}
\big ( 3 k_0^4 -4 k_0^2 m^2 +4 m^4+\Delta (8 k_0 m^2-12 k_0^3)
\nonumber \\ & + & \Delta ^2 (16 k_0^2 -8 m^2)-8 \Delta^3 k_0
+4 \Delta ^4 \big ) \nonumber \\
& - & \frac {\Delta (k_0^2 -4 m^2)^{3/2}}{2 k_0 (k_0^2-4 \Delta^2)^2}
(3 k_0^2- 8 m^2 -4 \Delta ^2) \Big ]
\end{eqnarray}
In the above, $g_A$ is the axial vector coupling constant chosen to
be $g_A=1.26$ \cite{chiral}. The $\rho NN $ coupling is as fitted
from the $NN$ scattering data \cite{grein} and is given as
$g_{\rho N N}$=2.6. We might note that, by definition,
the $\rho NN$ coupling used here is half of the coupling in
Ref \cite{chiral}.

For high energies, the $\rho N$ scattering process, $\rho N \rightarrow
\omega N$ also becomes important \cite{chiral}. 
The corresponding scattering amplitude is given as
\be
Im T_{\rho N}^{(\pi \omega)}
= \frac  {g_A^2 g_{\omega \rho \pi}^2}{48 \pi f_\pi^2 m_\pi^2}
\frac {k_0^2  {\vec { k_\omega}}^3 \Big [2 {\vec { k_\omega}}^2 
{m_N^* }^2 - m_N^* (k_{\omega 0} -k_0)\big [(k_{\omega 0}-k_0)^2 
-{\vec {k_\omega}}^2 \big ] \Big ]}{m_N^* (m_N^* +k_0)
(m_\pi ^2 -{m_\omega^*}^2+ 2 k_{\omega 0} k_0 -k_0^2)^2}
\label{piomg}
\ee
The interaction Lagrangian describing $\omega \rho \pi$ as used above,
is given as \cite{bali}
\be
{\cal L}_{\omega \rho \pi}=
\frac {g_{\omega \pi \rho}}{m_\pi}\epsilon _{\mu \nu \alpha  \beta}
\partial ^\mu \omega^\nu \partial ^\alpha {\rho}^{\beta}_i
{\pi_i}
\label{lwrpi}
\ee
For the $\omega \rho \pi$ coupling we take the value 
$g_{\omega \rho \pi}$ =2 according to Ref. \cite{sourav}
consistent with the decay width of $\omega \rightarrow \pi \gamma$. 
The $\omega \rho \pi$ coupling as here is
similar to that used in \cite{weisezp} after accounting 
for a factor of $m_\pi/f_\pi$ difference in the definitions.
In the above, $|{\vec {k_\omega}}|
=\sqrt {\lambda (m_N^* +k_0,m_\omega^*,m_N^*)}/(2(m_N^*+k_0)$
and $k_{\omega 0}=\sqrt {{m_\omega^* }^2 +{\vec {k_\omega}}^2}$.
The K\"allen function $\lambda$ is defined by
$\lambda (x,y,z)= (x^2 -(y+z)^2)(x^2-(y-z)^2)$.
The present calculations for the collisional decay width of
$\rho$ meson take into account the mass
modifications of the nucleon and vector mesons ($\rho$ and $\omega$).
For $\Delta$, we assume that the mass scales in the same way as the
nucleon mass in the medium, with $m_\Delta$=1232 MeV as
the mass in vacuum.
It may be noted that the contributions given by (\ref{pinn}), (\ref{piddel})
and (\ref{piomg}) to the $\rho N$ scattering amplitude
correspond to the zero temperature situation. 
The broadening of the $\rho$ spectral function 
is significantly due to the finite density effects, 
which are expected to dominate over the temperature effects
for the SPS conditions \cite{fireball}. 

The contribution to the decay width of $\rho$ due to $\rho N$ scattering
processes, arising from the $\pi N$, $\pi \Delta$
and $\omega \pi$ loops are given in terms of the imaginary parts 
of the scattering amplitudes as \cite{chiral}
\be
{\Gamma_{\rho N}}^{coll} = \frac {\rho_B (Im T_{\rho N}^{(\pi N)}
+Im T_{\rho N}^{(\pi \Delta)}
+Im T_{\rho N}^{(\omega \pi)})}{k_0}
\ee
We might note here that the effect of $\rho N$ scattering on the
real part of the scattering amplitude has been seen to be
negligible \cite{chiral} and has not been considered here.


To calculate the decay width for the $\omega$-meson, we
write down the effective Lagrangian for the $\omega$ meson as
\cite {sakurai,bali}
\begin{equation}
{\cal L}_\omega=-\frac {em_\omega^2}{g_\omega}\omega^\mu A_\mu
+ \frac {g_{\omega 3 \pi}}{m_\pi ^3}\epsilon _{\mu \nu \alpha  \beta}
\epsilon _{ijk}\omega ^\mu \partial ^\nu \pi ^i  \partial ^\alpha \pi ^j  
\partial ^\beta \pi ^k. 
\label{lomg}
\end {equation} 
In the above, the first term refers to the direct coupling of
the vector meson $\omega$ to the photon,  and hence to the dilepton
pairs, as given by the vector dominance model.
The decay width of the $\omega$-meson in vacuum is dominated
by the channel $\omega \rightarrow 3 \pi$.
In the medium, the decay width for $\omega \rightarrow 3\pi$
is given as
\begin{eqnarray}
\Gamma _{\omega \rightarrow 3 \pi} & = & \frac {(2\pi)^4}{2 k_0}
\int d^3 {\tilde p}_1   d^3 {\tilde p}_2   d^3 {\tilde p}_3\,
\delta ^{(4)} (P-p_1-p_2-p_3) |M_{fi}|^2
\nonumber \\ && \Big [(1+f(E_1))(1+f(E_2))(1+f(E_3))
-f(E_1)f(E_2)f(E_3) \Big ],
\end{eqnarray}
where $d^3 {\tilde p}_i=\frac {d^3 p_i}{(2\pi)^3 2 E_i}$, $p_i$ and
$E_i$'s are 4-momenta and energies for the pions, and $f(E_i)$'s
are their thermal distributions. The matrix element $M_{fi}$ has 
contributions from the channels 
$\omega \rightarrow \rho \pi \rightarrow 3\pi$ (described by Eq.
(\ref{lwrpi}) and the direct decay $\omega \rightarrow 3\pi$
resulting from the contact interaction (second term in (\ref {lomg}))
\cite{bali,weisezp,kaymak}. After fitting $g_{\omega \rho \pi}$
from the $\omega \rightarrow \pi \gamma$ decay width, 
the point interaction coupling $g_ {\omega 3 \pi}$ is determined
by fitting the partial decay width $\omega \rightarrow 3 \pi$ in vacuum
(7.49 MeV) to be 0.24 \cite{dlp}. The contribution arising from the 
direct decay is up to around 15 \%, which is comparable to the results of 
\cite{bali,weisezp}. 

With the modifications of the vector meson masses in the hot and dense 
medium, the contribution from the decay mode
$\omega \rightarrow \rho \pi$ becomes accessible \cite{dlp,sourav} 
when $m^*_\omega > m^*_\rho +m_\pi$. This is also considered 
in the present work. There may also be collisional broadening effects \cite{chiral} 
for the $\omega$ meson at high energies due to baryons.
However, these are seen sensitive to the meson-baryon form factors
\cite{chiral} and have not been considered in the present calculations. 

\section{Parametrisation of the photon self energy in the hadronic
phase}

Lattice simulations indicate that QCD undergoes a phase transition 
at a critical temperature, $T_c$, above which there is QGP phase, 
with quarks and gluons as the relevant degrees of freedom. 
As one approaches the critical temperature from the above, the quarks and 
gluons get confined to form the hadrons, which become the effective 
degrees of freedom for temperatures below $T_c$. In this section,
we discuss the parametrisation of the photon self energy in the
present scenario for the hadronic phase. The photon here
couples to the vector mesons ($\rho$ and $\omega$) and 
the electromagnetic current-current correlator can be related 
to the currents generated by these mesons which are calculated 
using a Lagrangian describing the hadronic phase.
The averaged photon spectral function is defined as \cite{fireball} 
\begin{equation}
R (q)=\frac {12 \pi}{q^2} {\rm {Im}} \bar \Pi (q),
\end{equation}
where the photon self energy is related to the sum of the vector meson 
self energies as
\begin{equation}
{\rm Im} \bar \Pi (q)=-\frac {1}{3} \sum _{V} 
{\Pi ^\mu _\mu} ^{V} (q).
\end{equation}
In the limit of $\vec q \rightarrow 0$, the averaged photon spectral
function can be written as
\begin{equation}
R (q)={ 12 \pi}{q_0^2} \sum _{V} 
{\rm Im} {\Pi} ^{V} (q_0)
\equiv 12  \pi \sum _{V} 
{\rm Im} {\tilde \Pi} ^{V} (q_0).
\end{equation}

In the Walecka model, the vector self energies can be parametrised
as a Breit-Wigner distribution with an energy dependent decay width,
along with a continuum as \cite{jane}
\begin{equation}
{\rm Im}{\tilde \Pi} ^V (q_0)=f_V^2 \frac {q_0 \Gamma_V (q_0)}{
(q_0^2 -m_V ^2)^2 +(q_0 \Gamma_V (q_0))^2}
+ \frac {C_V}{8\pi} \Big (1+\frac {\alpha_s (q_0)}{\pi} \Big )
\frac {1}{1+e^{(q_0^V -q_0)/\delta}}
\end{equation}
for the vector mesons, $V=\rho,\omega$. In the above, $f_\rho$ and 
$f_\omega$ are the coupling constants corresponding to the vector- 
meson-photon interactions. The continuum part is described as a smooth
function in the energy \cite{shuryak,shuryak1} instead of a 
step function as is usually adopted in QCD sumrule calculations
\cite{hatlee,chiral}. The values, $C_\rho$=1 and $C_\omega$=1/9
give the correct asymptotic limits for the spectral functions,
$\lim_{q_0 \rightarrow \infty} R_\rho$ =3/2 and 
$\lim_{q_0 \rightarrow \infty} R_\omega$ =1/6
\cite{shuryak,shuryak1,hatlee,chiral}.
Also, the energy dependence of the strong coupling constant occurring
in the meson self energies has been taken into consideration and is given
as $\alpha_s (q_0)=0.7/\ln(q_0/0.2)$ \cite{shuryak}, where $q_0$ is 
in units of GeV. The values for the parameters $q_0^{\rho,\omega}$
correspond to the threshold energies above which asymptotic freedom is 
restored and quark model estimates for cross-sections become valid.
The vacuum values for the parameters appearing in the continuum part
of the spectral function as derived from the experimental data
of $e^+ e^- \rightarrow hadrons$ are $q_0^\rho=1.3$ GeV,
$q_0^\omega=1.1$ GeV, and $\delta=0.2$ GeV \cite{shuryak}.
The vacuum values for the other parameters in the Breit-Wigner
part of the photon spectral function are $f_\rho$=152 MeV,
$f_\omega$=50 MeV,
$m_\rho$=770 MeV, $m_\omega$=783 MeV, $\Gamma_\rho$=151 MeV,  
$\Gamma_\omega$ =7.5 MeV. The masses of the vector
mesons are replaced by the medium modified masses 
including quantum correction effects \cite{dlp}, 
as a simple pole approximation \cite{hatlee,jane1}.
However, the energy dependence of the decay widths 
is considered. We assume the medium modified threshold energies,
${q_0 ^V}^*$ to be given by the simple scaling
\cite{jane}
$\frac {{q_0^V}^*}{q_0^V}=\frac {m_V^*}{m_V}$.
The above parametrisation of the photon spectral function for the
hadronic phase is used for studying the dynamical evolution of the strongly 
interacting matter arising in a ultrarelativistic heavy ion collision
in an expanding fireball model \cite{fireball}. In the following section,
we briefly discuss the parametrisation of the photon self energy 
in the QGP phase described by a quasiparticle picture \cite{quasi}.

\section{Photon self energy in the QGP phase}

The QGP phase is described \cite{quasi} using a quasiparticle picture.
The model treats quarks and gluons as massive thermal quasiparticles
with their properties determined so as to be compatible with the lattice QCD 
data. For temperatures much larger than $T_c$, the thermal masses can
be calculated in the hard thermal loop (HTL) approximation using the
perturbative QCD techniques. For temperatures near the phase transition,
however, the coupling $\alpha_s$ becomes large thus invalidating any
perturbative techniques. Around $T_c$, a power law fall--off is assumed
for the thermal masses, based on the conjecture that the phase 
transition is either weakly first order or second order as indicated 
by the lattice calculations. The thermodynamic quantities for the QGP 
are calculated in terms of two functions $B(T)$ and $C(T)$ introduced 
in the model,
which account for the thermal vacuum energy and the onset of confinement 
for temperature approaching $T_c$. The quasiparticles are by construction
noninteracting and hence the quasiparticle $q\bar q$ becomes the only
contribution to the photon self energy \cite{quasi}.
In the QGP phase the time like photon couples to the continuum
of thermally excited $q \bar q$ states and subsequently converts
into a dilepton pair, which gives the contribution to the dilepton emission
rate from the QGP phase.

\section{The fireball evolution model}

\subsection{Expansion and flow}

Our fundamental assumption is to treat the fireball matter as thermalized
from an initial proper time scale $\tau_0$ until breakup time $\tau_f$. For simplicity,
we assume a spatially homogeneous distribution of matter. Since some volume elements 
move with relativistic velocities, it is sensible
to choose volumes corresponding to a given proper time $\tau$ for the calculation of
thermodynamics, hence the thermodynamic parameters temperature $T$, entropy density
$s$, pressure $p$, chemical potentials $\mu_i$ and energy density $\epsilon$ become
functions of $\tau$ only for such a system. In the following, we refer to $\tau$ as the
time measured in a frame co-moving with a given volume element.

In order to make use of the information coming from lattice QCD
calculations, we proceed by calculating the thermodynamical response to
a volume expansion that is parametrized in such a way as to reproduce the experimental
information about the flow pattern and HBT correlations as closely as possible.
As a further simplification, we assume the volume to be cylindrically symmetric around
the beam (z)-axis. Thus, the volume is characterized by the longitudinal extension
$L(\tau)$ and the transverse radius $R(\tau)$ and we find
\begin{equation}
\label{E-Volume1}
V(\tau) = \pi L(\tau) R^2(\tau).
\end{equation}

In order to account for collective flow effects, we boost individual volume
elements according to a position-dependent velocity field. For the transverse
flow, we make the ansatz
\begin{equation}
\eta_T(r, \tau) = r/R_{rms}(\tau)\eta_T^{rms}(\tau)
\end{equation}
where $R_{rms}(\tau)$ denotes the
root mean square radius of the fireball at $\tau$ and $\eta_T^{rms}(\tau)$ the
transverse rapidity at $R_{rms}$.

For the longitudinal dynamics, we start with the experimentally measured width of the rapidity 
interval of observed hadrons $2\eta_f^{front}$ at breakup. From this, we compute the longitudinal velocity of the
fireball front at kinetic freeze-out
$v_f^{front}$. We do not require the initial expansion velocity $v_0^{front}$ to coincide
with $v_f^{front}$ but instead allow for a longitudinally accelerated expansion. 
This implies that during the evolution $\eta = \eta_s$ is not valid (with $\eta_s$ the spacetime 
rapidity $\eta_s = 1/2 \ln ((t+z)/(t-z))$ unlike in the non-accelerated case.

The requirement that the acceleration should be a function of $\tau$ and
that the system stays spatially homogeneous for all $\tau$ determines
the velocity field uniquely if the motion of the front is specified.
We solve the resulting system of equations numerically \cite{Synopsis}.
We find that for not too large rapidities $\eta < 4$ and accelerations
volume elements approximately fall on curves $const. = \sqrt{t^2 - z^2 }$ 
and that the flow pattern can be approximated
by a linear relationship between rapidity $\eta$ and  spacetime rapidity $\eta_s$ as
 $\eta(\eta_s) = \zeta\eta_s$
where $\zeta = \eta^{front}/\eta_s^{front}$
and $\eta^{front}$ is the rapidity of the cylinder front. 
In this case, the longitudinal
extension can be found calculating the invariant volume $V = \int d\sigma_\mu u^\mu$ as
\begin{equation}
L(\tau) \approx 2 \tau \frac{\text{sinh }((\zeta -1) \eta_s^{front}(\tau))}{(\zeta -1)} 
\end{equation}
with $\eta_s^{front}(\tau)$
the spacetime rapidity
of the cylinder front. This is an approximate
generalization of the boost-invariant relation $L(\tau) = 2 \eta^{front} \tau$ which can be derived
for non-accelerated motion.

\subsection{Parameters of the expansion}

In order to proceed, we have to specify the longitudinal acceleration $a_z(\tau)$ (which in
turn is used to calculate $\eta_s^{front}(\tau)$ numerically), the
initial front velocity $v_0^{front}$ and the expansion pattern of the radius $R(\tau)$
in proper time.

In principle, one would require $a = \nabla p/\epsilon$. However, our model
framework contains a homogeneous distribution of matter, therefore 
$\nabla p=0$ everywhere except at the surface
of the cylinder. In order to keep this approximation but nevertheless use a
more realistic acceleration, we make the additional assumption that in a realistic
situation a drop in temperature would leave the shape of the pressure
distribution rather unchanged while reducing the overall magnitude. With this
assumption, $\nabla p \sim c\cdot p$. Therefore, we make the ansatz
\begin{equation}
a_z = c_z \cdot \frac{p(\tau)}{\epsilon(\tau)}
\end{equation}
which allows a soft point in the EoS where the ratio $p/\epsilon$ gets small
to influence the acceleration pattern. $c_z$ and $v_0^{front}$ are 
model parameters governing the longitudinal expansion and fit to data.

Since typically longitudinal expansion is characterized by larger
velocities than transverse expansion, i.e. $v_z^{front} \gg v_T^{front}$, we
treat the radial expansion non-relativistically. We assume that the radius of the
cylinder can be written as 
\begin{equation}
R(\tau) = R_0 + c_T \int_{\tau_0}^\tau d \tau'  \int_{\tau_0}^{\tau'} d \tau'' \frac{p(\tau'')}{\epsilon(\tau'')}
\end{equation}

The initial radius $R_0$ is taken from overlap calculations. This leaves a parameter
$c_T$ determining the strength of transverse acceleration which is also fit to
data. The final parameter characterizing the expansion is its endpoint given by $\tau_f$,
the breakup proper time of the system. 

\subsection{Thermodynamics}

We assume that entropy is conserved throughout the thermalized expansion phase. Therefore, 
we start by fixing the entropy per baryon from the number of produced
particles per unit rapidity (see e.g. \cite{ENTROPY-BARYON}). 
Calculating the number of participant baryons (see \cite{fireball})
we find the total entropy $S_0$. The entropy density at a given proper time
is then determined by $s=S_0/V(\tau)$.

We describe the EoS in the partonic phase by a quasiparticle
interpretation of lattice data which has been shown to reproduce lattice
results both at vanishing baryochemical potential $\mu_B$  and
finite $\mu_B$ \cite{quasi} (see these references for details of the model).

For the phase transition temperature, we choose $T_C = 170$ MeV
based on lattice QCD computations at finite temperature for case of
two light and one heavy quark flavour \cite{lat2}. We also
note that no large latent heat is observed in the transition and
model the actual thermodynamics as a crossover rather than a sharp
phase transition. Nevertheless, in the calculation we assume quarks and gluons as degrees
of freedom above $T_C$ and hadrons below to simplify computations. Since the
time the system spends in the vicinity of the transition temperature is
small compared with the total time for dilepton emission however, any
error we make by this assumption is bound to be small as soon as we
consider the measured rates which represent an integral over the time
evolution of the system folded with the emission rate.

Since a computation of thermodynamic properties of a strongly interacting
hadron gas close to $T_C$ is a very difficult task, we follow a simplified approach
in the following:
We calculate thermodynamic properties of the hadron
gas at kinetic decoupling where interactions cease to be important. Here, we 
have reason to expect that an ideal gas will be a good description and calculate the EoS
with the help of an ideal resonance gas model. Using the framework of statistical hadronization
\cite{Hadrochemistry}, we determine the overpopulation of pion phase space
by pions from decays of heavy resonances created at $T_C$ and include this
contribution (which gives rise to a pion-chemical potential of order $\mu_\pi \approx 120$ MeV
into the calculation.
We then choose a smooth interpolation between decopling temperature
$T_f$ and transition temperature $T_C$ to the EoS
obtained in the quasiparticle description. This is described in greater detail in
\cite{fireball}.

With the help of the EoS and $s(\tau)$, we are now in a position to compute
the parameters $p(\tau), \epsilon(\tau), T(\tau)$ as well.
Since the ratio $p(\tau)/\epsilon(\tau)$ appear in the expansion
parametrization, we have to solve the model self-consistently.

\subsection{Solving the model}

In order to adjust the model parameters, we compare with data on transverse momentum
spectra and HBT correlation measurements. 
This is discussed in greater detail in \cite{HBT, Synopsis}.

In \cite{FREEZE-OUT}, a very similar model is fit to a large set of experimental data,
providing different sets of parameters $T_f, v_{\perp f}, R_f, \eta_f^{front}$.
Although we use a different (box vs. Gaussian) longitudinal distribution of matter,
we use the parameters from this analysis as a guideline for our transverse dynamics
where this difference should not show up and determine $\eta_f^{front}$ separately.
Specifically, we use the set {\bf b1} from  \cite{FREEZE-OUT} for the transverse
dynamics.

By requiring $R(\tau_f) = R_f$ and $v_T^{front} = v_{\perp f}$ we can determine the
model parameters $c_T$ and $\tau_f$. $c_z$ is fixed by the requirement
$\eta^{front} (\tau_f) = \eta_f^{front}$. The remaining parameter $v_0^{front}$
now determines the volume (and hence temperature) at freeze-out and can be adjusted such 
that $T(\tau_f) = T_f$.

The model for 5\% central 158 AGeV Pb-Pb collisions at SPS is characterized by
the following scales: Initial long. expansion velocity $v_0^{front} = 0.5c$, thermalization
time $\tau_0 = 1$ fm/c, 
initial temperature $T_0 = 305$ MeV, duration of the QGP phase $\tau_{QGP} = 7$ fm/c,
duration of the hadronic phase $\tau_{had} = 9$ fm/c, total lifetime $\tau_f - \tau_0
= 16$ fm/c,  r.m.s radius at freeze-out $R_f^{rms} = 8.55$ fm, transverse expansion
velocity $v_{\perp f} = 0.537 c$. 

For the discussion of dileptons, we require the fireball evolution for other
than 5\% central collisions. In this case, we make use of simple scaling
arguments based on the initial overlap geometry and the number of collision
participants. For a detailed description, see \cite{fireball, Charm}.

In \cite{fireball}, it has been shown that this scenario is able to
describe the measured spectrum of low mass dileptons, and
in \cite{Hadrochemistry} it has been demonstrated that under the assumption
of statistical hadronization at the phase transition temperature $T_C$,
the measured multiplicities of hadron species can be reproduced.
In \cite{Charm}, the model has been shown to describe
charmonium suppression correctly. None of these quantities is, 
however, very sensitive to the detailed choice of the equilibration 
time $\tau_0$. Therefore, we have only considered
the `canonical' choice $\tau_0 =$ 1 fm/c so far. The calculation 
of photon emission within the present framework provides the opportunity 
to test this assumption and to limit the choice of $\tau_0$.
In \cite{Photons}, this has been investigated in some detail.
Within the present framework, the limits $0.5$ fm/c $< \tau_0 < $ $3$ fm/c
could be found. Variations within these limits, however, do not
affect the spectrum of dileptons with invariant mass below 1 GeV 
significantly.

\section{Dilepton Emission}

The emission of dileptons in the model is calculated using the differential emission rate
\begin{equation}
\frac{dN}{d^4 x d^4q}  =  \frac{\alpha^2}{\pi^3 q^2} \ \frac{1}{e^{\beta q^0}
- 1} \ \mbox{Im}\bar{\Pi}(q, T) = \frac{\alpha^2}{12\pi^4} \frac{R(q, T)}{e^{\beta q^0}
- 1}  ,\label{dileptonrates}
\end{equation}
where $\alpha = e^2/4\pi$ and we have neglected the lepton
masses. Eq.(\ref{dileptonrates}) is valid to order $\alpha$ in the electromagnetic
interaction and to all orders in the strong interaction. Its main ingredient is the
 temperature-dependent spectral function $R(q, T)$. 

The differential rate of eq.(\ref{dileptonrates}) is integrated over the space-time history of the collision to
compare the calculated dilepton rates with the CERES/NA45 data \cite{expt}
taken in Pb-Au collisions at 158 AGeV (corresponding to
a c.m. energy of $\sqrt{s} \sim 17$ AGeV) and 40 AGeV ($\sqrt{s} \sim 8$ AGeV).  The CERES experiment is a
fixed-target experiment. In the lab frame, the CERES
detector covers the limited rapidity interval $\eta = 2.1-2.65$, {\em i.e.} $\Delta\eta = 0.55$. 
We integrate the calculated rates over the
transverse momentum $p_T$ and average over $\eta$, given that $d^4p = M p_T \
dM  \ d\eta \ dp_T \ d\theta.$
The formula for the space-time- and $p$-integrated dilepton rate hence becomes
\begin{equation}
\frac{d^2N}{dM d\eta} =  \frac{2\pi M}{\Delta \eta} \int \limits_{\tau_0}^{\tau_{f}}
d\tau \  \int  d\eta \ V(\eta,T(\tau))\int
\limits_0^\infty dp_T \ p_T
 \ \frac{dN(T(\tau),M, \eta,
p_T)}{d^4 x d^4p} \ Acc(M, \eta, p_T), \label{integratedrates}
\end{equation}
where $\tau_{f}$ is the freeze-out proper time of
the collision, $V(\eta,T(\tau))$ describes the proper time evolution of 
volume elements moving at different rapidities 
and the function $Acc(M, \eta, p_T)$ accounts for the
experimental acceptance cuts specific to the detector. At the CERES
experiment, each electron/positron track is required to have a transverse
momentum $p_T > 0.2$ GeV, to fall into the rapidity interval $2.1 < \eta <
2.65$ in the lab frame and to have a pair opening angle $\Theta_{ee} > 35$
mrad. 
Finally, for comparison with the CERES data, the resulting rate 
is divided by $dN_{ch}/d\eta$, the rapidity density of charged particles.

As in \cite{fireball}, we also include the effects of the overpopulation of the
pion phase space due to decay processes of heavy resonances to the emission from 
hadronic matter by introducing a
temperature-dependent pion chemical potential $\mu_\pi(T)$.

In addition to the thermal emission of dileptons, we also consider dileptons
from vacuum decays of vector mesons after the thermal decoupling of the fireball
and hard dileptons from initial Drell-Yan processes. For details of the calculation
of these contributions see ref.~\cite{fireball}.
 
A final remark: The spectral function considered in the present work is only available
below 1 GeV invariant mass range. Therefore, we do not include thermal emission
of dileptons from hadronic matter above that scale when discussing the Walecka model
results. This leads to small differences between the two approaches for large invariant
masses. This region, however, is not dominantly filled by emission from the hadronic phase
but by the QGP and Drell-Yan contribution.  

\section{Results and Discussions}


\begin{figure}
\begin{center}
\includegraphics[width=8cm, angle=-90]{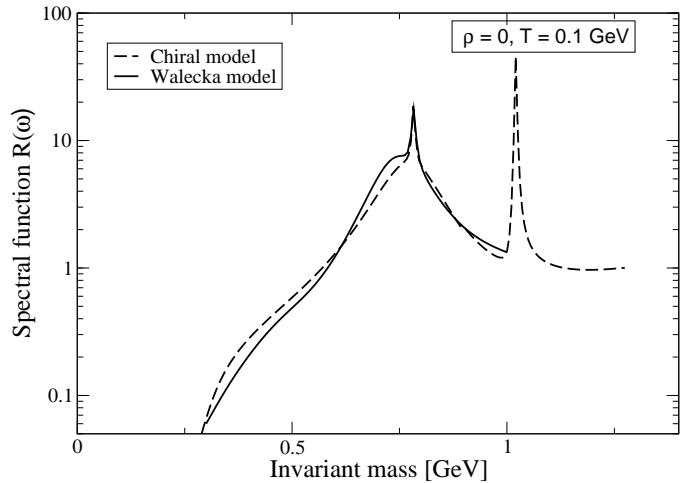}
\caption{Photon spectral function at T=100 MeV.}
\label{spectt100}
\end{center}
\end{figure}

We now discuss the results obtained for the photon spectral function 
in the hadronic matter arising due to medium modification 
of the vector mesons with the quantum correction effects 
from baryons and scalar mesons, and its effect on the dilepton emission spectra. 
Contrary to the mean field approximation, there is large drop of the vector meson masses 
due to vacuum polarisations from the nucleon sector. This shifts the $\rho$ and $\omega$ peaks 
in the dilepton spectra to lower values \cite{dlp,souravhat}. 
The additional quantum correction effects from the sigma mesons
lead to considerable increase of the $\omega$ decay width \cite{dlp}. 

In the present investigation, we choose the renormalised sigma meson self interaction
coupling $\lambda_R$ to be 5, corresponding to the value of incompressibility
of nuclear matter to be 329 MeV \cite{dlp}. As stated earlier, 
the dilepton spectra are studied in a mixed scenario of QGP and hadronic matter 
using the fireball model as described in the previous section. 
The photon spectral function as obtained in the
present hadronic scenario is compared with the earlier calculation
obtained in a chiral SU(3) model. This is described in \cite{chiral} for finite 
density and \cite{chiral-T} for finite temperature modifications.
In \cite{fireball}, we assumed that it is possible to factorize these contributions.

The spectral functions
in both the models are shown in figure \ref{spectt100} 
for vanishing baryon density 
and a temperature of 100 MeV. As expected due to absence 
of any pronounced medium effects for this temperature, 
the two models yield similar results. 

\begin{figure}
\begin{center}
\parbox[b]{8cm}{
\includegraphics[width=6.7cm, angle=-90]{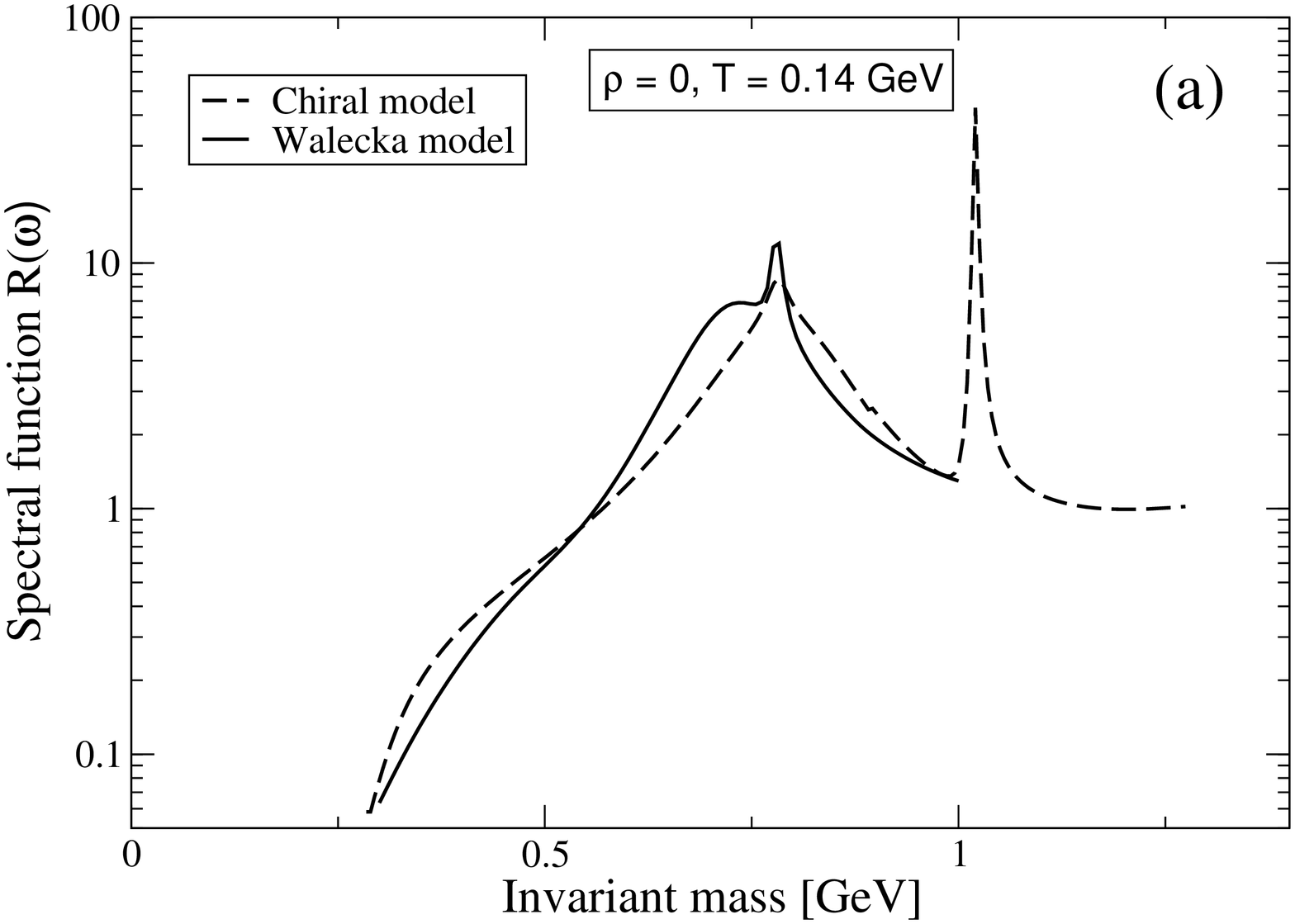}}
\parbox[b]{8cm}{
\includegraphics[width=6.7cm, angle=-90]{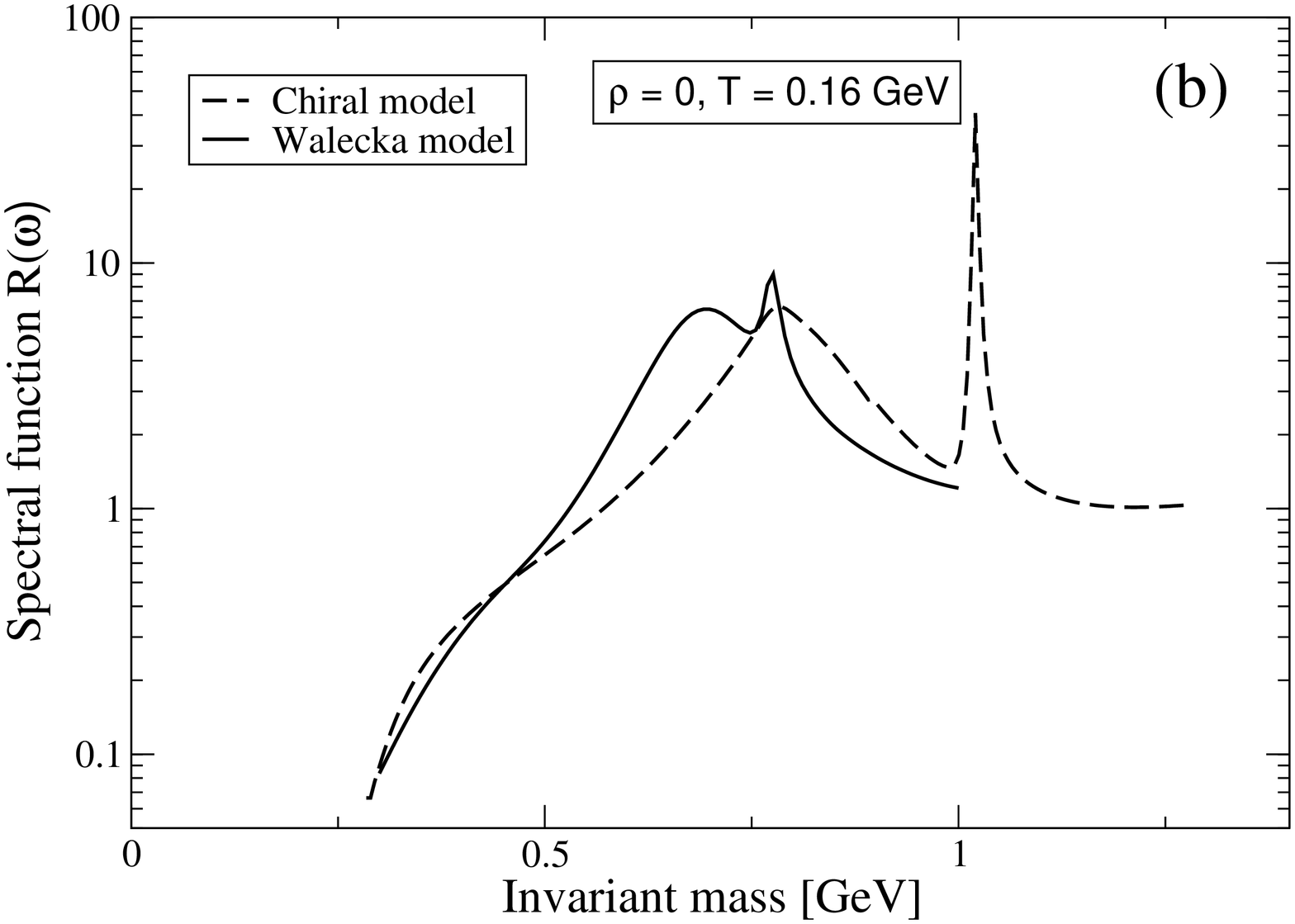}}
\caption{
Photon spectral function at T=140 MeV and T=160 MeV.}
\label{specttemp}
\end{center}
\end{figure}

\begin{figure}
\begin{center}
\parbox[b]{8cm}{
\includegraphics[width=6.7cm, angle=-90]{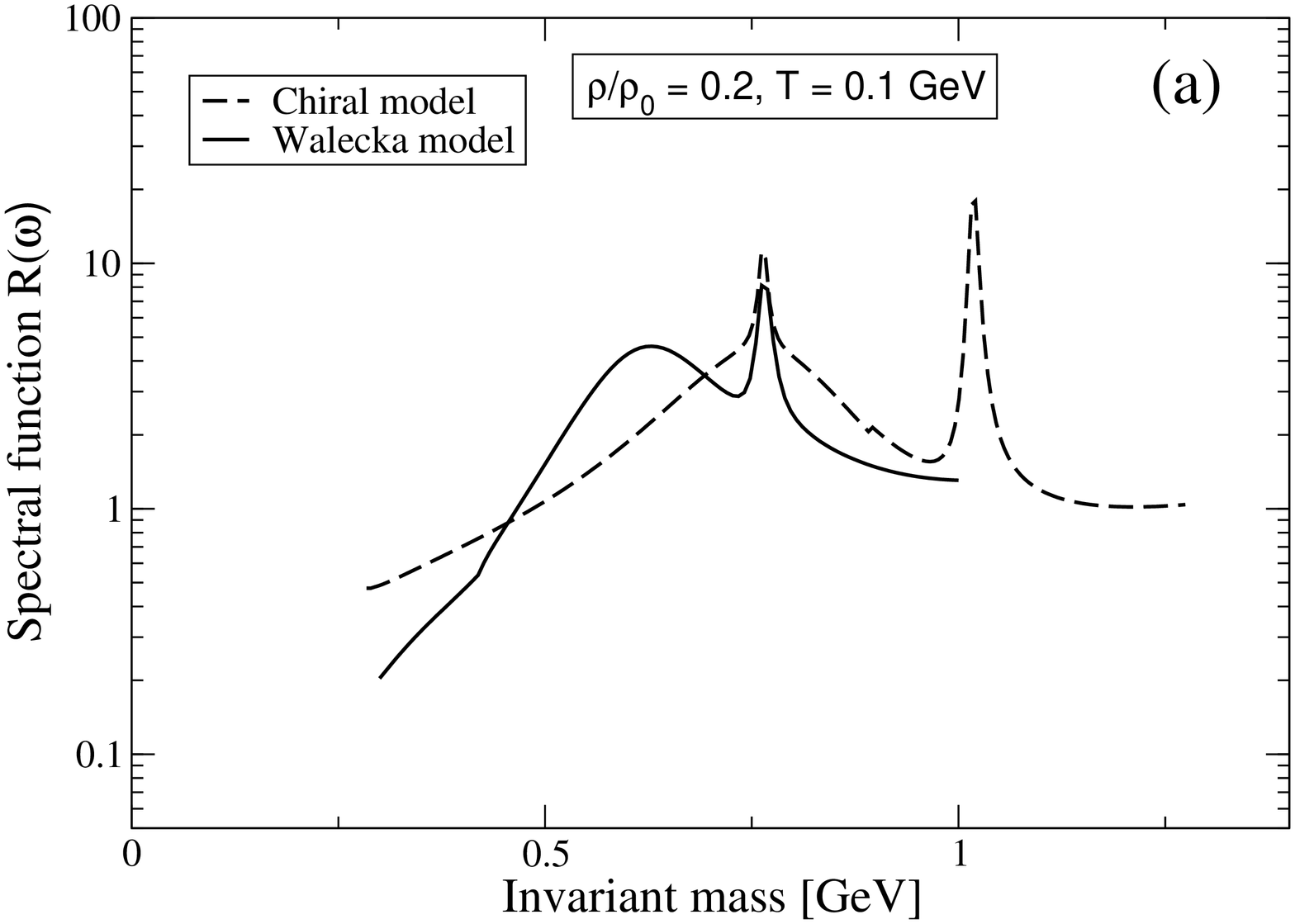}}
\parbox[b]{8cm}{
\includegraphics[width=6.7cm, angle=-90]{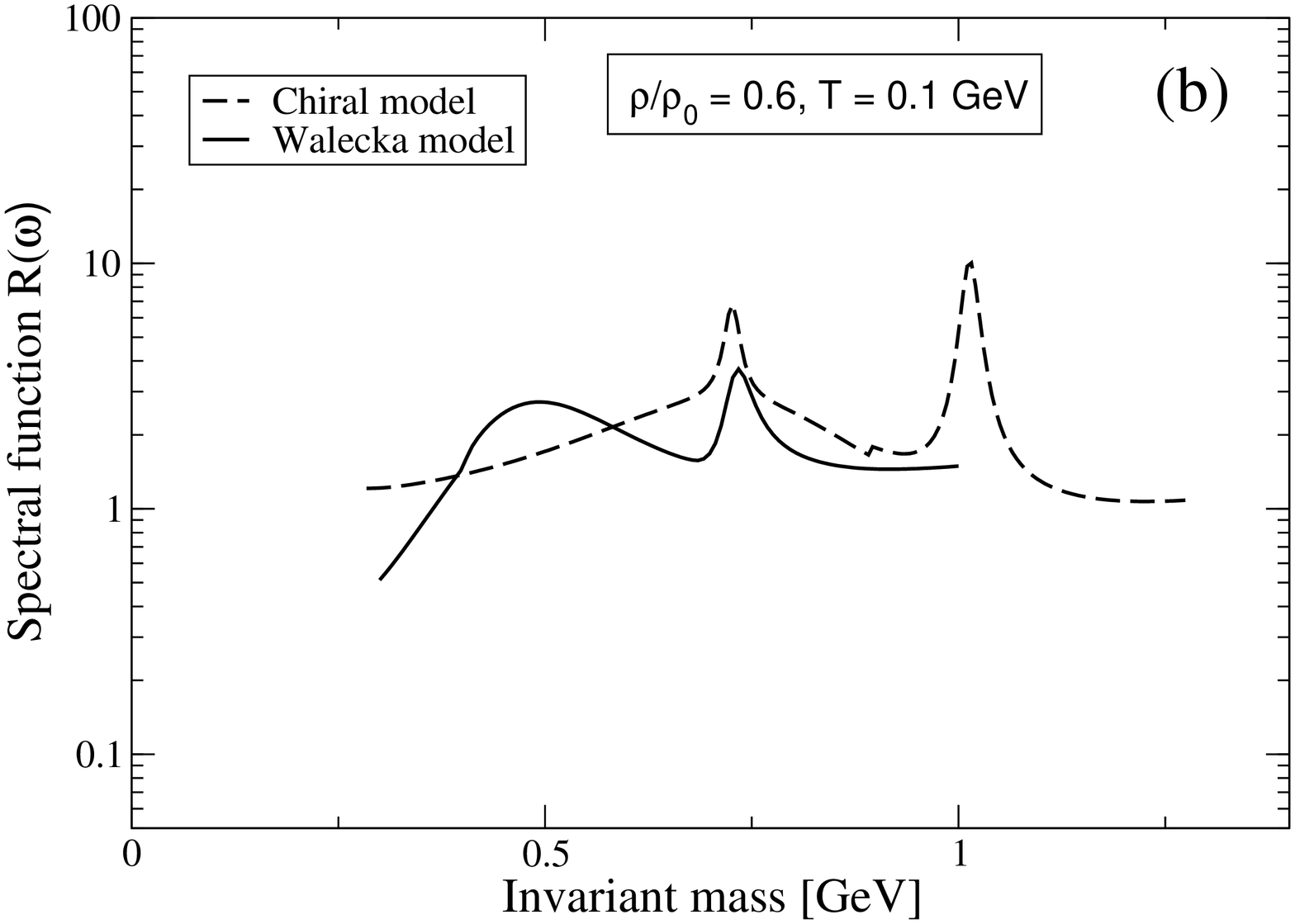}}
\caption{
Photon spectral function at T=100 MeV and at densities 0.2$\rho_0$
and 0.6$\rho_0$, with $\rho_0$=0.17 $\rm {fm}^{-3}$.}
\label{spectdens}
\end{center}
\end{figure}

In figure \ref{specttemp}, the temperature dependence of the photon
spectral function is shown. The spectral function in the present hadronic 
scenario is seen to develop a distinct $\rho$ peak at higher temperatures.
This is a reflection of the fact that with quantum correction effects, 
the $\rho$-mass, due to the tensorial coupling, has a larger drop 
as compared to the mass of the $\omega$-meson. 
In the present calculations,
the effect of temperature on the $\rho$-mass 
is rather moderate up to around a temperature of 160 MeV or so.
This is in line with the previous calculation
of \cite{gale}. The broadening of the $\rho$ peak at higher temperature
takes place as the Bose enhancement dominates over the 
effect of drop in $\rho$-mass, leading to an increase in the
$\rho$ decay width. On the other hand, the density dependence of the
$\rho$ mass is seen to be quite significant \cite{dlp} which leads to
the $\rho$ peak position shifted to lower values in the 
present calculation as illustrated in figure \ref{spectdens}.
The mass of the $\rho$-meson remains almost unchanged in the
chiral model, whereas the $\rho$- decay width has appreciable
enhancement in the medium due to inelastic processes, 
like $\rho N \rightarrow \pi N$, $\rho N \rightarrow \pi \Delta$, 
and also, from $\rho N \rightarrow \omega N$ at higher energies.
This basically leads to the $\rho$ peak to be completely dissolved
in the chiral model \cite{chiral}, leaving only a broad continuum.
In the absence of collisional effects, the decay width for $\rho$ 
arises from the process $\rho \rightarrow \pi\pi$. Here, 
the Bose-Einstein factors
in (\ref{gmrho}) have the effect of increasing the width in the
thermal medium, whereas the stronger dropping of the $\rho$-mass
in the medium at higher densities in the present calculations
has the opposite effect of decreasing the decay width and overcompensates 
the effect from the Bose enhancement factor. 
The collisional effects lead to considerable flattening
of the spectral function for the $\rho$ peak.


\begin{figure}
\includegraphics[width=9cm, angle=-90]{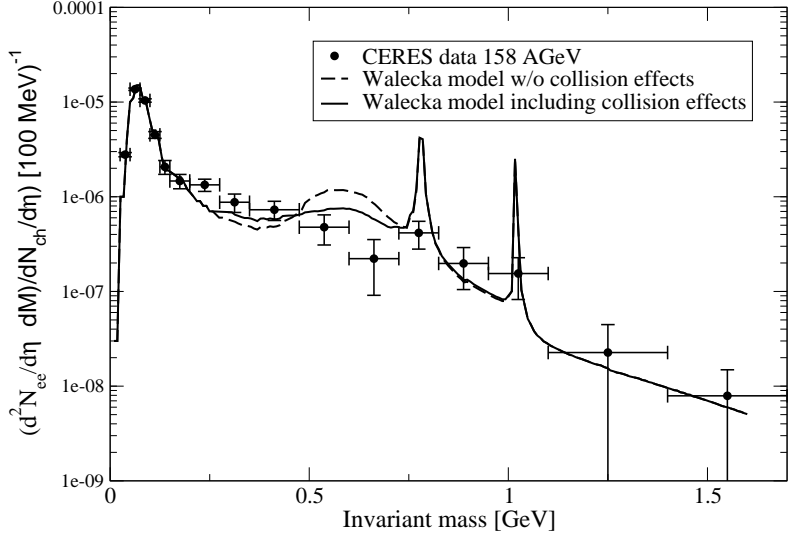}
\caption{
\label{comp158}
Dilepton emission rate for 158 AGeV at SPS for the Walecka model
with and without the collision effects.}
\end{figure}

\begin{figure}
\includegraphics[width=9cm, angle=-90]{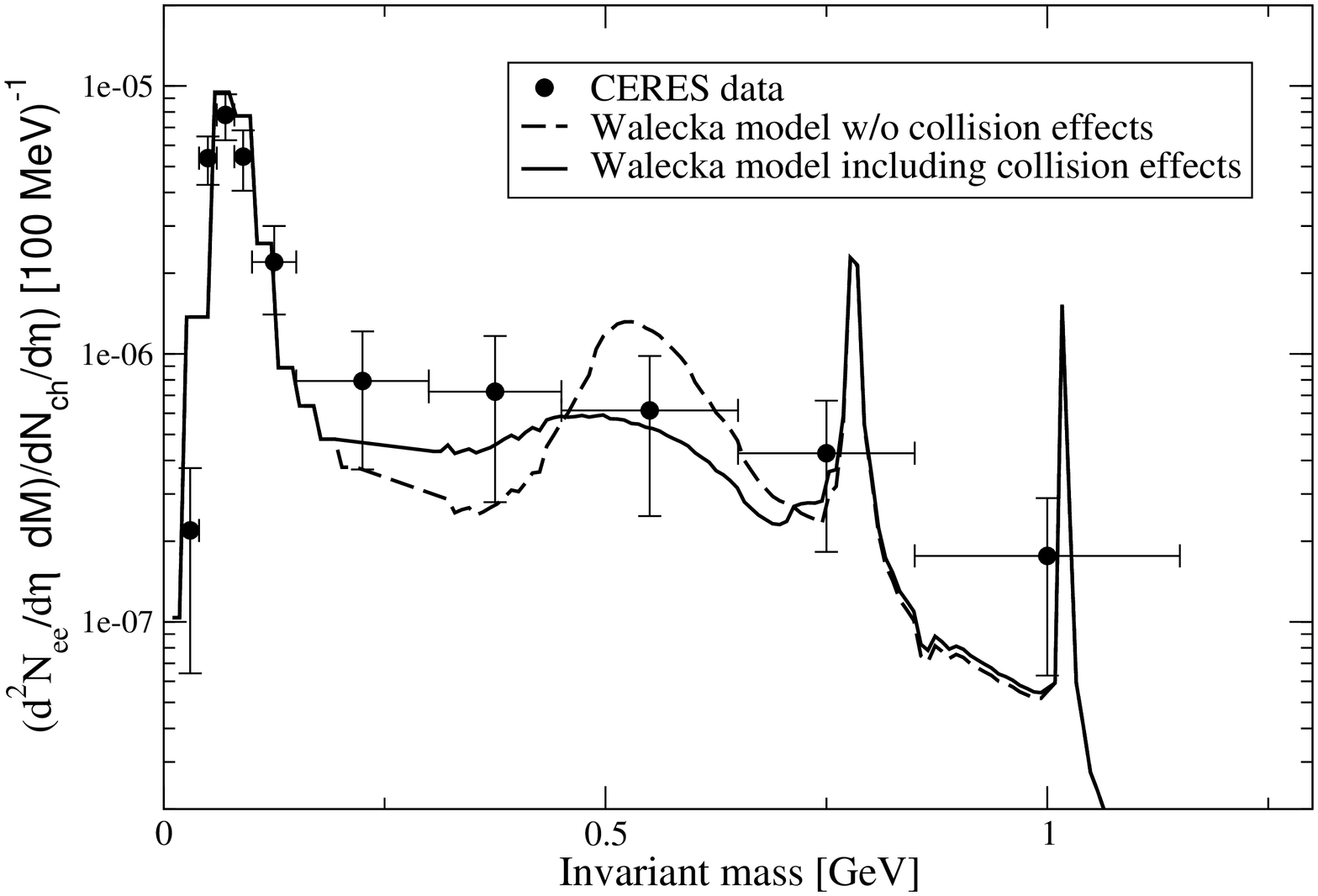}
\caption{
\label{comp40}
Dilepton emission rate for 40 AGeV at SPS for the Walecka model
with and without the collision effects.}
\end{figure}

\newpage
\begin{figure}
\begin{center}
\includegraphics[width=9cm, angle=-90]{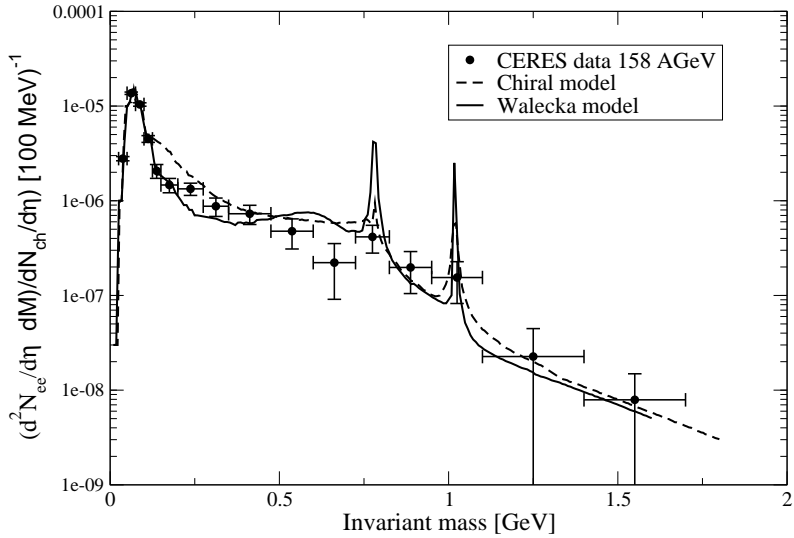}
\caption{
\label{dlp158}
Dilepton emission rate for 158 AGeV at SPS for the chiral 
and Walecka models.}
\end{center}
\end{figure}

Figures \ref{comp158} and \ref{comp40} show the dilepton spectra
using the fireball model in the present hadronic scenario
and a quasiparticle picture for QGP. The inclusion of collisional
effects due to scattering off by the nucleons is seen
to lead to considerable broadening of the $\rho$-peak in 
the dilepton spectra. The difference is seen to be more
prominent for SPS 40 AGeV corresponding to a higher baryon density.
The present hadronic scenario leads to a distinct structure 
in the dilepton emission rate due to the undissolved $\rho$ peak
in the absence of the contribution from the $\rho N$ scattering 
processes, which is seen to be flattened due to these collision
processes. The dropping of the vector meson masses in QHD in the 
relativistic Hartree approximation \cite{souravhat}
also has been seen to yield similar results.
In the absence of collisional processes, the drop in the $\rho$ mass 
leads to enhancement in the dilepton
yield below the vacuum $\rho$ mass, but does not
have the rather flat structure as observed in the data.
Inclusion of inelastic processes $\rho N \rightarrow \pi N$,
$\rho N \rightarrow \pi \Delta$, $\rho N \rightarrow \omega N$,
give rise to considerable broadening of the spectra below 
the vacuum $\rho$-mass.

\begin{figure}
\begin{center}
\includegraphics[width=9cm, angle=-90]{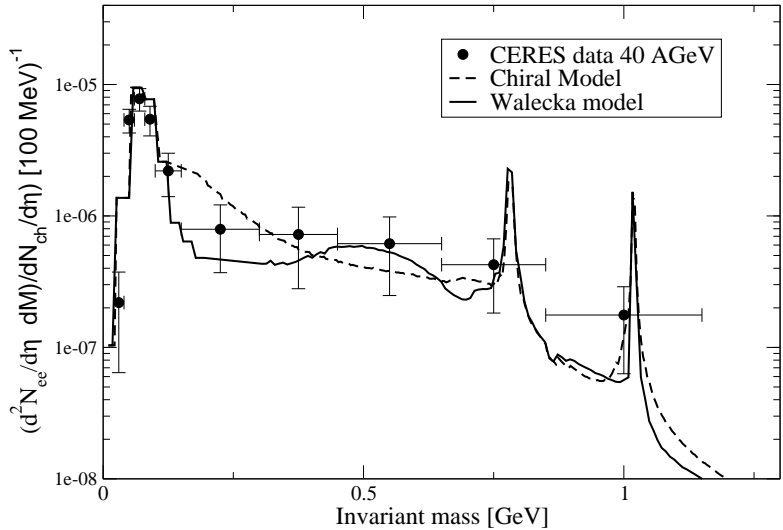}
\caption{
\label{dlp40}
Dilepton emission rate for 40 AGeV at SPS for the chiral 
and Walecka models.}
\end{center}
\end{figure}

Figures \ref{dlp158} and \ref{dlp40} compare the dilepton emission rates 
of the present investigation to the results of chiral
model description for hadronic matter as well as to
the results from SPS, 158 AGeV and SPS, 40 AGeV 30\% 
central Pb+Au collision.
It may be noted that the fireball evolution, detector acceptance,
contributions from the QGP and Drell-Yan background were modelled
exactly in the same way for both theoretical curves. 
The previous hadronic scenario based on a chiral model 
\cite{chiral, chiral-T} has the observed dissolved $\rho$ peak arising due to the 
Bose enhancement factors as well as the inelastic channels.
In the present calculations, the collisions by nucleons 
lead to significant broadening of the $\rho$-peak in
the dilepton spectra.

Contrary to what has been done in \cite{fireball}, we have not folded the
results with the finite energy resolution of the detector in order to
indicate what could be observed if the energy resolution were increased. 
The sharp peaks of $\omega$ and $\phi$ seen in the plots (absent in the plots in \cite{fireball}
and seemingly in contradiction with the data)
represent vacuum decays of these mesons left after kinetic decoupling of the fireball.
As apparent from \cite{fireball}, if the strength contained in these peaks is
smeared out to represent finite detector resolution, agreement with the data is achieved.

In \cite{Huovinen}, dilepton and photon emission are studied using a more detailed
description of the fireball matter using coarse-grained UrQMD and hydrodynamics. The
essential scales of the expansion dynamics however appear similar to our approach -
we also see dilepton emission for matter with average temperatures in the region
between 120 and 250 MeV for a time period of about 15 fm/c. This is no
surprise since essential scales of the evolution are dictated by the hadronic momentum
spectra (and HBT radii) to which our model is adjusted. We might therefore expect that
simplifying assumptions (such as introducing an average $T$ for given $\tau$) do not
play a large role when one integrates over the complete four volume of the expansion.

It seems that the main difference which might account for the fact that our approach
describes the data well between 0.3 and 0.6 GeV invariant mass whereas
\cite{Huovinen} is somewhat on the low side is the use of different
spectral functions. However, in order to test this a calculation in our
framework using yet other spectral functions needs to be carried out which is
beyond the scope of the present paper.

In \cite{RappReport}, different approaches to describe the dilepton
data for 158 AGeV beams are compared. Unfortunately, each of the
calculations shown employs a different fireball model and a different description
of in-medium modification of the vector mesons, emphasizing the need for studies
along the line of the present paper. However, the emerging picture for
BUU, transport and hydrodynamical calculations seems to be that to first
approximation, the normalization of the dilepton invariant mass spectrum
is determined by the fireball evolution, which is in turn strongly
constrained by the hadronic momentum spectra and 2-particle correlations
and should lead to similar essential scales of the fireball once the
builtup of a pion chemical potential due to the decay of heavy resonances created at
$T_C$ is taken into account. The shape of the spectrum is then dictated
by the spectral function. Here, a scenario employing a dropping in-medium
$\rho$ mass is shown to lead to similar results as broadening of the $\rho$
peak due to the fact that a $\rho$ mass dropping as a function of $T$ sweeps through
a large invariant mass region as $T$ changes throughout the fireball evolution.
In most approaches however, baryons seem to be crucial for the description of the spectral
shape, in line with our own findings.

Unfortunately, the relations of the thermal fireball descriptions investiagted in
\cite{RappReport} to essential
evolution scales is less clear --- for example, no comment is made how transverse
and longitudinal flow are implemented and what the relation of these models 
to the actual hadronic spectra at freeze-out would be and to what degree the
models are tuned to reproduce dilepton data. Nevertheless, the resulting
shape of the dilepton spectrum is expected to be close to a more detailed
calculation whereas the normalization can be expected to be less certain.

Based on the present data, no clear distinction can be made between different
approaches to calculate the photon spectral function at finite temperature and density.
However, choosing a model which can be shown to describe other hadronic observables
as well and using different spectral functions within one evolution model
seems crucial to make progress in understanding the differences between the approaches
on a more quantitative level.

\section{summary}
To summarize, we have investigated here the medium properties
of the vector mesons in the Walecka model including the 
vacuum polarization effects and their effects on the
dilepton spectra using a fireball model. The density dependence
of the vector meson properties are seen to be the dominant
medium effect as compared to the temperature dependence.  
We have considered a mixed scenario of QGP and hadronic matter \cite{fireball} 
and have compared the dilepton emission rates of the present calculations
to those from a previous calculation using a chiral SU(3) model, 
as well as to the experimental results from SPS at 158 AGeV and 40 AGeV.

Since the evolution of the fireball was not in any way
adjusted to the dilepton data but rather fixed from different
observables, we are in a position to compare the different spectral
functions with the data without being subject to large uncertainties 
regarding the medium evolution. 

Inclusion of vacuum polarisation effects in the Walecka model leads to 
dropping of these masses in the medium, contrary to the mean field approximation.
This gives rise to the shift of the $\rho$ and $\omega$ peaks in the dilepton spectra
to smaller values \cite{dlp,souravhat}. Due to the stronger medium modification, 
the $\rho$ peak is seen to be distinct from the $\omega$ peak in the spectra. 
When the scattering due to nucleons is not taken into account
in the present calculations, there is a more pronounced difference 
in the two models for SPS, 40 AGeV, as compared to SPS, 158 AGeV 
due to the higher baryon density. However, there is seen to be significant 
broadening of the spectra due to these collision effects within the present
investigation.

\begin{acknowledgements}
One of the authors (AM) would like to thank J. Reinhardt and S. Schramm 
for fruitful discussions. AM is grateful to the Institut f\"ur
Theoretische Physik for warm hospitality and acknowledges financial
support from Bundesministerium f\"ur Bildung und Forschung (BMBF).
TR would like to thank W.~Weise and R.~A.~Schneider for their support of
this work. Financial support was given in part by BMBF and GSI.
\end{acknowledgements}

\end{document}